\documentclass[a4paper,11pt]{article}
\usepackage{pos}

\title{The distribution amplitude of the $\eta_c$ meson}
\ShortTitle{Charmonia distribution amplitudes}

\author*[a]{Miguel Teseo San Jos\'{e} P\'{e}rez}
\author[a]{Beno\^{i}t Blossier}
\author[b]{Mariane Mangin-Brinet}
\author[c]{Jos\'{e} Manuel Morgado Ch\'{a}vez}

\affiliation[a]{Laboratoire de Physique des 2 Infinis Irène Joliot-Curie, CNRS/IN2P3,\\
  Université Paris-Saclay, F-91405 Orsay Cedex, France}

\affiliation[b]{Laboratoire de Physique Subatomique et de Cosmologie, CNRS/IN2P3,\\
  38026 Grenoble, France}

\affiliation[c]{D\'{e}partement de Physique Nucl\'{e}aire, Irfu/CEA-Saclay,\\
  91191 Gif-sur-Yvette Cedex, France}

\emailAdd{san-jose-perez@ijclab.in2p3.fr}
\emailAdd{blossier@ijclab.in2p3.fr}
\emailAdd{mariane@lpsc.in2p3.fr}
\emailAdd{jose-manuel.morgadochavez@cea.fr}

\abstract{We report on the first lattice determination of the pseudoscalar meson $\Petac$ light-cone \acl{da}, using a set of three \acs{cls} $N_f=2$ ensembles at a pion mass $m_{\pi} \sim \qty{270}{\mega\eV}$ and lattice spacings $a \sim \qtylist{0.076;0.066;0.049}{\femto\metre}$. Employing \acl{sdf}, we extract the \acl{pda} on the lattice for Ioffe times $\nu \leq 4.5$, and the various lattice spacings allow us to take the continuum limit. We employ a basis of Jacobi polynomials to parametrize the distribution amplitude, which allows to express the matching to the pseudo distribution in closed form, and we observe a strong effect which we attribute to the heavy charm-quark mass.}

\FullConference{The 40th International Symposium on Lattice Field Theory (Lattice 2023)\\
July 31st - August 4th, 2023\\
Fermi National Accelerator Laboratory\\}

\begin{document}
\maketitle

\section*{Introduction}

The discovery of the $\PJpsi$ state at \acs{slac} and \acs{bnl} in 1974 corroborated the existence of the charm quark and validated the \acs{gim} mechanism, which explains the suppression of flavor-changing neutral currents in experiments. Ever since, more quarkonium states have been discovered composed of a $Q\bar{Q}$ pair of heavy quarks. However, their rate of production is poorly understood even today, and several models trying to capture the essential features of the problem have emerged. For example, the \acl{cem} is based on the principle of quark-hadron duality, and so it expects the production cross section of quarkonia and the $Q\bar{Q}$ pair to be directly connected below the open $Q$ threshold. It owes its name to the assumption that a number of soft gluon emissions occur after the $Q\bar{Q}$ creation, decorrelating the initial and final states. Quite in the opposite direction, the \acl{csm} assumes the initial $Q\bar{Q}$ pair is already a color singlet and carries the same spin as the final hadron, and assumes that no meaningful emission of gluons occurs. Finally, \acl{nrqcd} builds the physical quarkonium as a linear combination of Fock states ordered by $\alpha_s$ and the relative velocity $v_{Q}$. Being an effective theory, however, it introduces a number of new parameters. See \cite{Lansberg:2019adr} for an overview of these methods.

In our project, we study the \ac{da} of the pseudoscalar meson $\Petac$, which appears in processes like the Higgs boson decay to charmonium states \cite{Han:2022rwq}. Besides, we plan to extend the methodology underlined here to the $\PJpsi$ \ac{da}, which will serve to probe the charm Yukawa coupling at the \acl{hl-lhc}. We employ a set of \acs{cls} $N_f=2$ lattice simulations to compute the \acl{pda}, and we employ the \ac{sdf} method \cite{Radyushkin:2017cyf} to relate the latter to the light-cone \ac{da} in the continuum limit.

\section*{From Euclidean space to the light cone}

\begin{figure}
\centering
\begin{subfigure}[b]{0.49\textwidth}
\centering
\includegraphics[width=\textwidth]{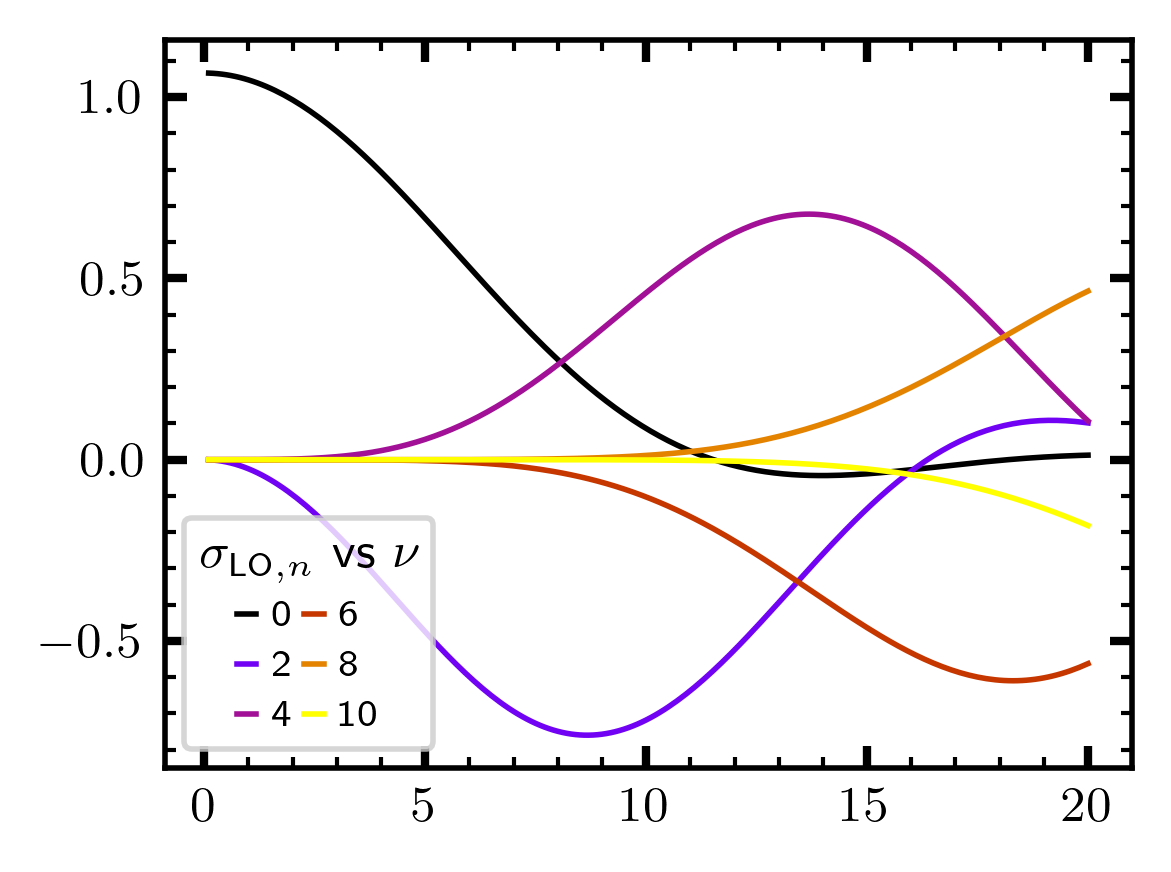}
\end{subfigure}
\begin{subfigure}[b]{0.49\textwidth}
\centering
\includegraphics[width=\textwidth]{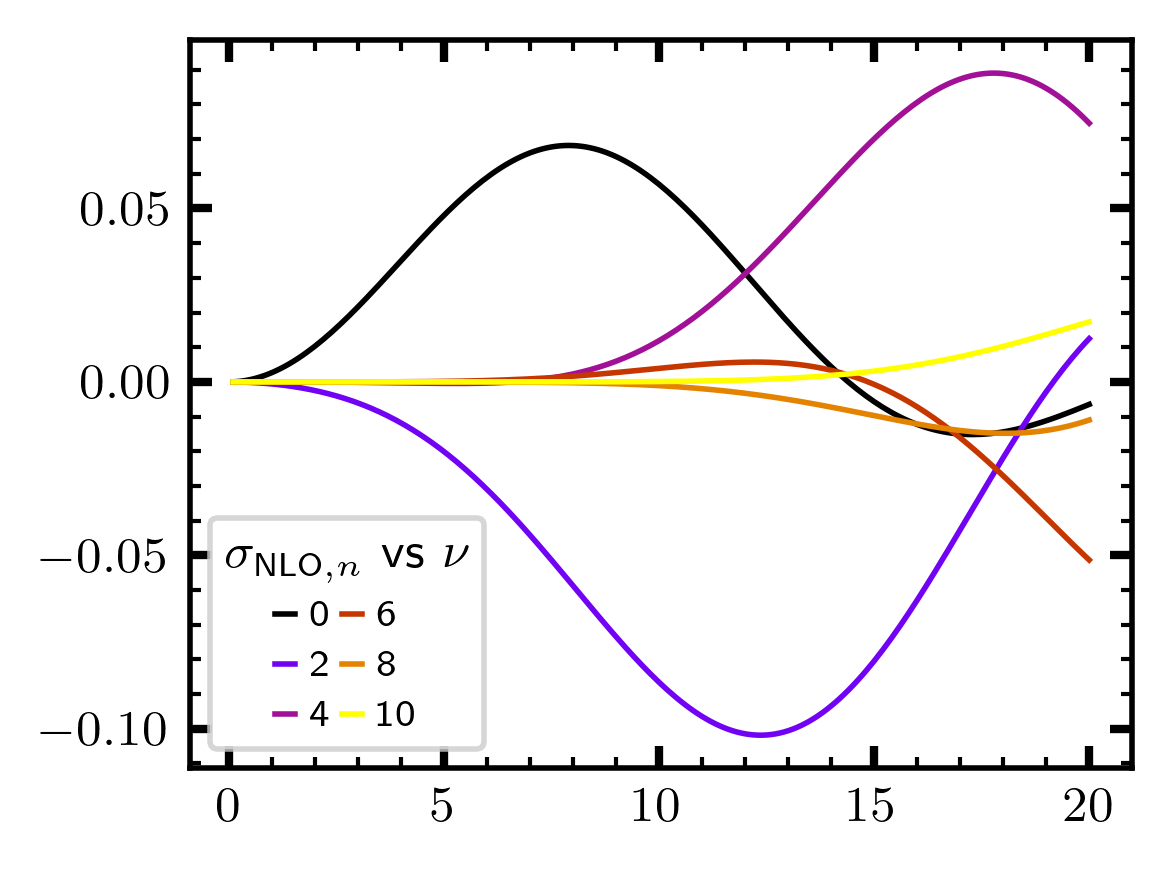}
\end{subfigure}
\caption{Left: The \acs{lo} contribution to $\sigma_{n}$ versus Ioffe time. Right: The \acs{nlo} contribution $\sigma_{\text{\acs{nlo}},n}=\sigma_n-\sigma_{\text{\acs{lo}},n}$ versus Ioffe time. As representative values, we use $\alpha=2$, $\abs{z}/a=5$, and $a=\qty{0.0658}{\femto\metre}$. Only the $n$-even coefficients are nonzero.}
\label{fig:sigma}
\end{figure}

Let us start establishing the relation between the light-cone \ac{da} and the \acl{pda} \cite{Ji:2013dva,Radyushkin:2017cyf,Orginos:2017kos} in the continuum limit. In Euclidean space, we compute the matrix element
\begin{equation}
\label{eq:mel}
M^{\mu}(p,z) = \mel{\Petac(1s)}{\APcharm(z)\gamma^{\mu}\gamma_5 W(z,0)\Pcharm(0)}{0},
\end{equation}
where $W$ is a Wilson line, $\bra*{\Petac(1s)}$ is the final pseudoscalar meson state, $\ket{0}$ is the QCD vacuum, and $\Pcharm$, $\APcharm$ are the charm-quark lines. The matrix element $M^{\mu}$ is evaluated at particular values of the momentum $p$ and the Wilson line extension $\abs{z}$.
We decompose $M^{\mu}$ in a purely $\textup{twist}>2$ contribution $\mathcal{M}^{\prime}$ and a term $\mathcal{M}$ which gives access to the leading-twist distribution amplitude \cite{Radyushkin:2017cyf},
\begin{equation}
M^{\mu}(p,z) = 2p^{\mu}\mathcal{M}(p,z)+z^{\mu}\mathcal{M}^{\prime}(p,z).
\end{equation}
To extract the leading-twist part of $\mathcal{M}$, we set the momentum $p^{\mu}=(0,0,p^3,E)$ and the equal-time separation $z^{\mu}=(0,0,z^3,0)$. Choosing $\mu=4$, we get rid of the main high-twist contamination $\mathcal{M}^{\prime}$. And to remove the prefactor $2E$ and obtain a renormalized quantity, we form the \acs{rgi} ratio \cite{Orginos:2017kos,Karpie:2018zaz}
\begin{equation}
\label{eq:rgi-ratio}
\phi(\nu,z) \equiv \dfrac{M^{4}(p,z)M^{4}(0,0)}{M^{4}(0,z)M^{4}(p,0)},
\end{equation}
where $\nu=pz$ is the Ioffe time. For \acp{da}, we can shift $\phi$ by a phase to obtain a real quantity, $\tilde{\phi}$, which we call the \ac{rpitd} and define as
\begin{equation}
\label{eq:rpitd}
\tilde{\phi}(\nu,z) = e^{-\iu\nu/2}\phi(\nu,z).
\end{equation}
The \ac{rpitd} is related to the $\msbar$ \ac{itd} $\tilde{\phi}(\nu,\mu)$ via \cite{Radyushkin:2019owq}
\begin{equation}
\label{eq:matching}
\tilde{\phi}(\nu,z) = \int_0^1 \dd{w} C(w,\nu,z\mu)~\tilde{\phi}(w\nu,\mu),
\end{equation}
where $C(w,\nu,z\mu)$ is a matching kernel known up to \acs{nlo} in $\alpha_s$ \cite{Radyushkin:2019owq}, and $\mu=\qty{3}{\giga\eV}$ is the $\msbar$ renormalization scale. Throughout these proceedings we use $N_f=2$ dynamical flavors, $N_c=3$ colors, $\Lambda_{\text{\acs{qcd}}}=\qty{330}{\mega\eV}$ \cite{FlavourLatticeAveragingGroupFLAG:2021npn}, and the aforementioned renormalization scale. Since the Ioffe time and the Bjorken $x$ are conjugate variables, one can reconstruct the light-cone \ac{da} $\phi(x,\mu)$ from the real Fourier transform
\begin{equation}
\label{eq:x2v}
\tilde{\phi}(\nu,\mu) = \int_0^1 \dd{x} \cos\left[\nu\left(x-1/2\right)\right] \phi(x,\mu).
\end{equation}
To extract the \ac{da} from a lattice calculation, one needs to apply \cref{eq:matching,eq:x2v} consecutively after taking the continuum limit. This requires evaluating $\tilde{\phi}(\nu,z)$ at each $(\nu,z)$ for several lattice spacings. Unfortunately, this is not practical or possible at the moment. Instead, we follow an approach already used in the study of \aclp{pdf} \cite{Karpie:2021pap}, and parametrize the light-cone \ac{da} in terms of a basis of Jacobi polynomials,
\begin{equation}
\label{eq:da-parametrization}
\phi(x,\mu)=(1-x)^{\alpha}x^{\alpha}\sum_{n=0}^{\infty} d_{2n}^{(\alpha)}\tilde{J}_{2n}^{(\alpha)}(x),
\end{equation}
where $\tilde{J}_{2n}^{(\alpha)}(x) \equiv J_{2n}^{(\alpha)}(2x-1)$ and $d_{2n}$ are free coefficients which should be determined by the data. Usually, one specifies a basis of Jacobi polynomials via two parameters, $\alpha$ and $\beta$, but in the case of charmonium \acsp{da} without electromagnetism, one expects $\phi(x,\mu)$ to be symmetric around $x=1/2$. Therefore, we set $\alpha=\beta$ and use only even $n$s, which have the correct symmetry. In this case, the Jacobi polynomials are proportional to Gegenbauer polynomials, which are commonly used in the study of \acp{da} \cite{Belitsky:2005qn}. Then, the relation between $\tilde{\phi}(\nu,z)$ and $\phi(x,\mu)$ given by \cref{eq:matching,eq:x2v} together can be expressed as a series with the coefficients $d_{2n}^{(\alpha)}$. The main strategy is to develop in a Taylor series the cosine in \cref{eq:x2v} and express the powers of $x-1/2$ using Jacobi polynomials. Employing the orthogonality relation of the latter, it is possible to find
\begin{equation}
\label{eq:leading-twist}
\tilde{\da}(\ioffet,z)=\sum_{n=0}^{\infty} d_{2n}^{(\alpha)}\sigma_{2n}^{(\alpha)}(\nu,z\mu).
\end{equation}
The functions $\sigma_n$ work as a basis in Ioffe time, just as the Jacobi polynomials in the Bjorken variable, but given their lengthy expression, we defer to write them down to a future publication. For the present work, let us suffix to say that one can divide $\sigma_n$ in a \acs{lo} piece $\sigma_{\text{\acs{lo}},n}$, setting to zero the strong coupling in $C(w,\nu,z\mu)$, and a \acs{nlo} part $\sigma_{\text{\acs{nlo}},n}=\sigma_n-\sigma_{\text{\acs{lo}},n}$. We plot them both in \cref{fig:sigma}. Looking closer at this figure, each $\sigma_n$ peaks at a different value of $\nu$, providing information on different regions. From the fact that $\tilde{\phi}(\nu=0,z) = 1$ and that $\sigma_0$ is the only nonzero function at $\nu=0$, we can determine that $d_0$ is a Beta function,
\begin{equation}
d_0^{(\alpha)}=\dfrac{1}{B(1/2,1+\alpha)}.
\end{equation}

\section*{The lattice calculation}

\begin{figure}
\centering
\begin{subfigure}[b]{0.49\textwidth}
\centering
\includegraphics[width=\textwidth]{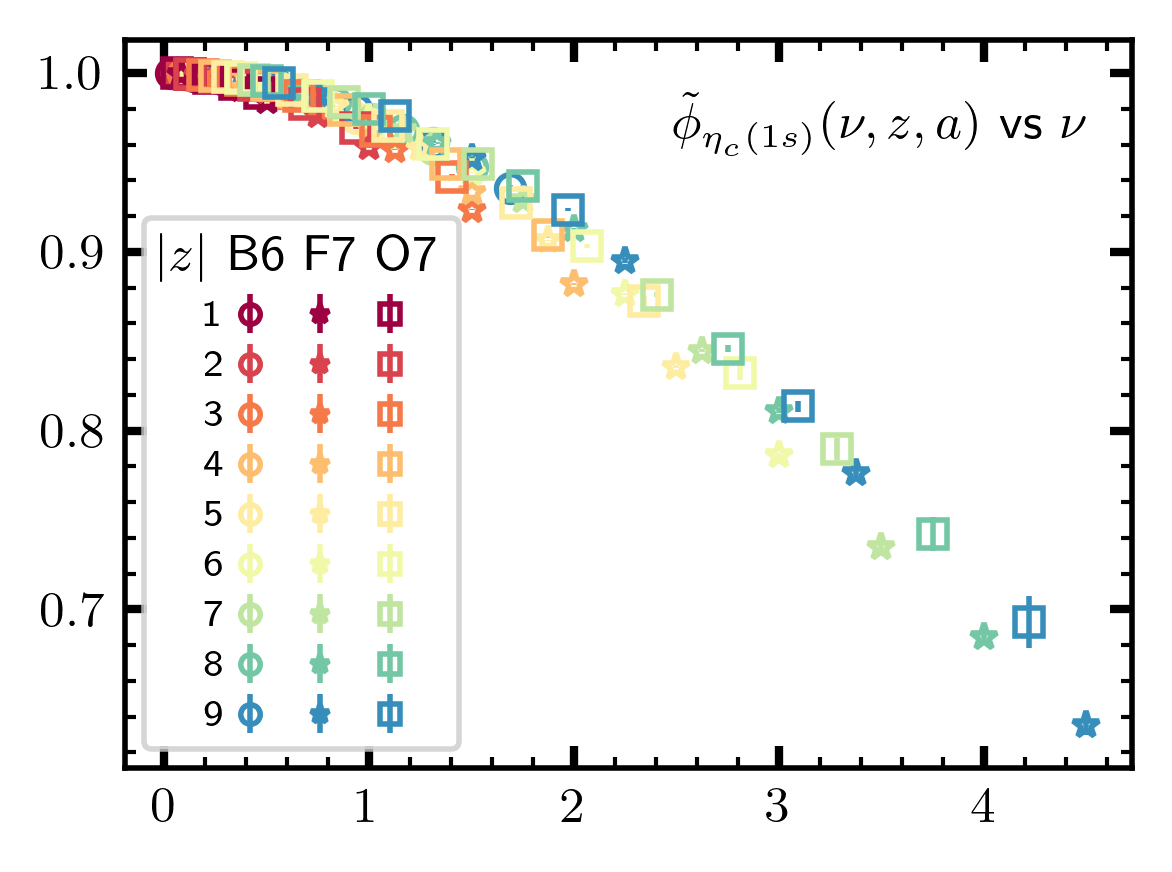}
\end{subfigure}
\begin{subfigure}[b]{0.49\textwidth}
\centering
\includegraphics[width=\textwidth]{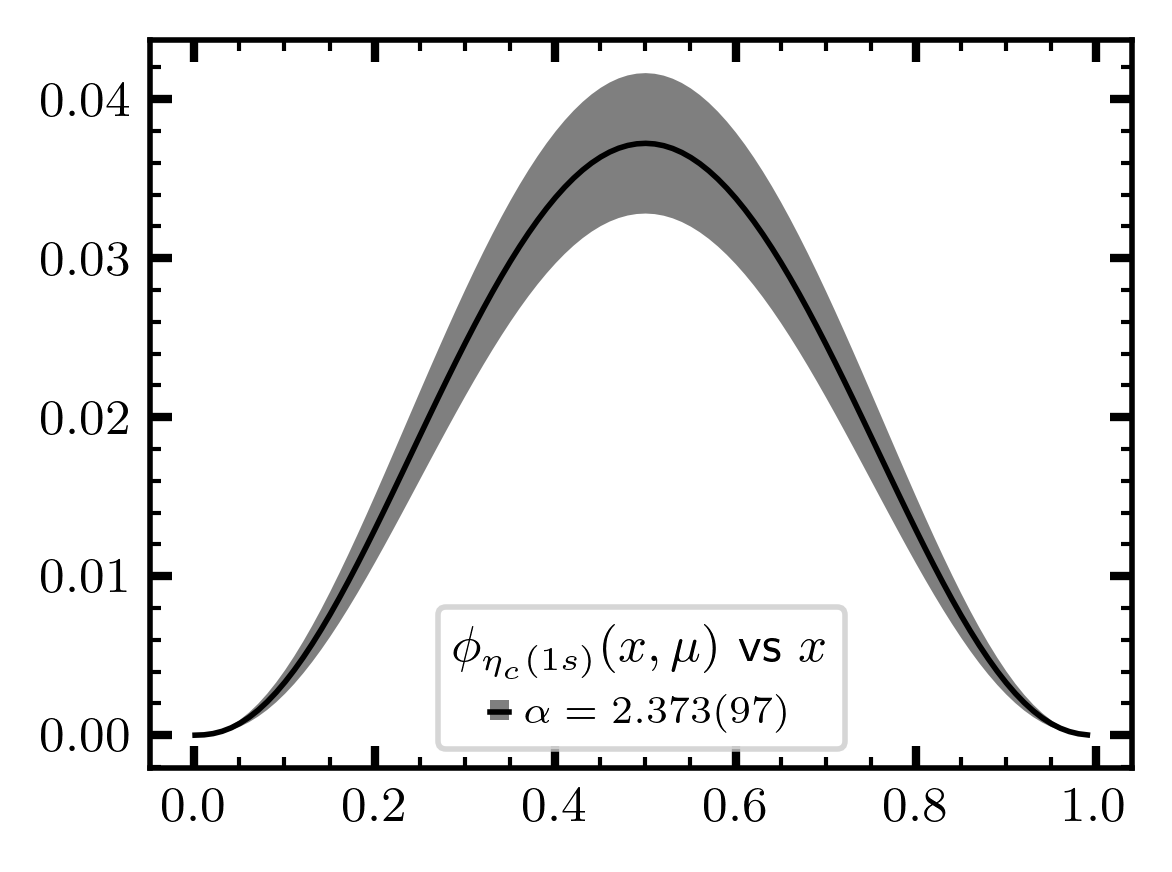}
\end{subfigure}
\caption{Left: The values of the \acl{rpitd} used for this study. The different colors indicate the length of the Wilson line, while the markers refer to the ensembles in \cref{tab:cls-ensembles-included}. Right: The light-cone \acl{da} as obtained from our extrapolation to the continuum.}
\label{fig:da}
\end{figure}

We employ a set of three $N_f=2$ \acs{cls} ensembles \cite{Fritzsch:2012wq,Heitger:2013oaa} with parameters and statistics gathered in \cref{tab:cls-ensembles-included}. Their action is formed by a Wilson plaquette in the gauge sector and two mass-degenerate flavors of non-perturbatively $\order{a}$ improved Wilson quarks. Their pion mass is $m_{\Ppi} \sim \qty{270}{\mega\eV}$, $\kappa_{\Pcharm}$ is fixed by the physical mass of the $\PDs$ meson, $m_{\PDs} = m_{\PDs,\text{phy}}=\qty{1969(1)}{\giga\eV}$ \cite{ParticleDataGroup:2022pth}, and the strange quark is set to its physical value. All ensembles employ the locally deflated Schwarz preconditioned \gls{gcr} solver, but B6 and F7 use the \gls{ddhmc} algorithm, while O7 employs \gls{mphmc}. The coupling constant was defined via the Schr\"{o}dinger functional, and the scale was determined using the kaon decay constant. Regarding the quark-connected charm propagators, we compute two sets of data: the matrix elements in \cref{eq:mel}, and pseudoscalar correlators for $\Petac$ states in the $A_1^{-+}$ cubic representation. Both sets include four different Gaussian smearings, six momenta computed via twisted boundary conditions, and ten $\group{U}{1}$ random sources defined in a random time slice, diluted in spin, and employing the one-end trick. For \cref{eq:mel} alone, we compute ten different Wilson-line extensions, and we employ the $\Petac$ correlators to solve a $4 \times 4$ \acs{gevp} and find the optimal interpolator to the pseudoscalar meson at each momentum.
\begin{table}
\centering
\begin{tabular}{ccccccccc}
\toprule
  id &
  $\beta$ &
  $a~[\unit{\femto\metre}]$ &
  $m_{\Ppi}~[\unit{\mega\eV}]$ &
  $m_{\Petac}~[\unit{\mega\eV}]$ &
  $\kappa_{\ell}$ &
  $\kappa_c$ &
  $\#$ cnfgs &
  $\#$ hits \\
\midrule
  B6 & 5.2 & 0.0755(9)(7) & 281 & 2929 & 0.13597 & 0.12529 & 118 & 10 \\
\midrule
  F7 & 5.3 & 0.0658(7)(7) & 265 & 2955 & 0.13638 & 0.12713 & 200 & 10 \\
\midrule
  O7 & 5.5 & 0.0486(4)(5) & 268 & 2972 & 0.13671 & 0.13022 & 164 & 10 \\
\bottomrule
\end{tabular}
\caption{CLS ensembles used on our analysis. The various columns indicate, from left to right,
the ensemble label, its bare strong coupling constant, the lattice spacing, the approximate $\Ppi$ and $\Petac$ mass, the light- and charm-quark $\kappa$, and the number of gauge configurations and hits.}
\label{tab:cls-ensembles-included}
\end{table}
On a lattice simulation, we start from $M^{\mu}(p,z,t,a,m_{\Ppi},m_{\Petac})$, where besides the momentum and Wilson line, there are dependencies in time, lattice spacing, and $\Ppi$ and $\Petac$ masses. The time dependence vanishes in the double ratio \cref{eq:rgi-ratio}, while the imaginary part of the matrix elements disappears due to the extra phase in \cref{eq:rpitd}. Then, we extract a real value for each $\tilde{\phi}(p,z,a,m_{\Ppi},m_{\Petac})$ fitting the time dependence to a constant minimizing a chi-square function. We repeat the fit for ensembles B6, F7, and O7 individually to obtain the plot on the \acs{lhs} of \cref{fig:da}, which shows the \ac{rpitd} with different markers for each ensemble and different colors for each Wilson line.

Finally, the continuum extrapolation involves not only removing the regulator, but also estimating the high-twist contamination and the impact of the small difference in the meson masses between the three ensembles. The basic piece of the extrapolation model is \cref{eq:leading-twist}, which intends to describe the leading-twist, continuum \ac{itd}. Given the interval of Ioffe times in our study, we observe that we are only sensitive to the $d_0$ coefficient, such that the \ac{da} at this stage is determined by the $\alpha$ parameter alone. Regarding the lattice artifacts, they start at $\order{a}$ because there is no Symanzik improvement available for the bilocal operators in \cref{eq:mel}. Besides, we can use CP symmetry to determine that $\tilde{\phi}(p,z)=\tilde{\phi}(-p,-z)$, such that odd powers of $a$ must be functions of $a\abs{p}$ or $a/\abs{z}$. Furthermore, together with Wilson-line-dependent lattice artifacts, there can also be global lattice artifacts $a\Lambda_{\text{\acs{qcd}}}$ ---note that we use $\Lambda_{\text{\acs{qcd}}}$ to render all quantities dimensionless. The lattice artifacts may also depend on the Ioffe time, and to model such dependence we can exploit the same basis of Jacobi polynomials used for the \ac{itd}. Following the example of \cite{Karpie:2021pap}, we define nuisance functions
\begin{equation}
A_r^{(\alpha)}(x)=(1-x)^{\alpha}x^{\alpha}\sum_{s=1}^{S_{a,r}} a_{r,2s}^{(\alpha)} \tilde{J}_{2s}^{(\alpha)}(x),
\end{equation}
where $a_{r,2s}^{(\alpha)}$ are the nuisance fit parameters. We are interested in the Ioffe time dependence,
\begin{equation}
  A_r^{(\alpha)}(\nu)
  = \int_0^1 \dd{x} A_r^{(\alpha)}(x) \cos\left[\left(x-\dfrac{1}{2}\right)\nu\right]
  = \sum_{s=1}^{S_{A,r}} a_{r,2s}^{(\alpha)}\sigma_{\text{LO},2s}^{(\alpha)}(\nu).
\end{equation}
Note that the term $s=0$ vanishes due to the condition $\tilde{\phi}(\nu=0,z)=1$. From examination, we consider it is enough to include $r=1$ only and set $S_{A,1}=1$, such that the nuisance function reduces to one single term with one fit parameter. In the fit, each term will have a similar nuisance function with its own free parameter. With regard to higher-twist effects, their leading behavior goes as $z^2\Lambda_{\text{\acs{qcd}}}^2$, and we include another nuisance function for them. Finally, the mass corrections are the least understood. Since we work with heavy quarks, it is possible that the charm-quark mass affects significantly not only the nuisance functions, but also the leading twist \ac{itd}. Unfortunately, no mass corrections are included in the calculation of the matching kernel $C(w,\nu,z\mu)$ \cite{Radyushkin:2019owq}. Some results do exists for other distributions but are not applicable here. They can only guide us to create an Ansatz, which is also informed by the fit quality of the extrapolation. We introduce an identical term for the pion masses, although for the latter we only wish to model the mistunings in the pion mass, not account for any physics. Then, the fit model we employ  at this stage is
\begin{equation}
  \label{eq:continuum-extrapolation-model}
  \begin{aligned}
    \tilde{\da}(\nu,z,a)
    =&
    ~\tilde{\da}_{\text{lt}}(\nu,z)
    + \dfrac{a}{\abs{z}}
      \left(A_1(\nu) + \log\left(\dfrac{m_{\eta_c}}{m_{\eta_c,\text{phy}}}\right)D_1(\nu)\right)
    \\
    &+ a\Lambda_{\text{\acs{qcd}}} B_1(\nu)x
    + z^2\Lambda_{\text{\acs{qcd}}}^2 C_1(\nu)
    + \dfrac{a}{\abs{z}} \log\left(\dfrac{m_{\Ppi}}{m_{\Ppi,\text{phy}}}\right)E_1(\nu).
  \end{aligned}
\end{equation}
Note that the product of $a/\abs{z}$ by the nuisance functions produce terms proportional to $a\abs{p}$ automatically, and that we normalize by the physical meson masses $m_{\Ppi,\text{phy}}=m_{\Ppizero}=\qty{134.9768(5)}{\mega\eV}$ and $m_{\Petac,\text{phy}} = \qty{2983.9(4)}{\mega\eV}$ \cite{ParticleDataGroup:2022pth}. The model includes one nonlinear fit parameter, $\alpha$, and five linear ones, $a_{1,2}$, $b_{1,2}$, $c_{1,2}$, $d_{1,2}$, and $e_{1,2}$. We leverage this asset and rewrite the linear parameters in terms of $\alpha$ following the \acl{varpro} algorithm \cite{doi:10.1137/0710036}, which helps to speed up and stabilize the chi-square minimization. The model \cref{eq:continuum-extrapolation-model} fits our data with $\chi^2/\text{dof}=165/134=1.23$. We find $\alpha = \num{2.37 \pm 0.10}$, $a_{1,2} = \num{4.216 \pm 0.026}$, $b_{1,2} = \num{2.53 \pm 0.18}$, $c_{1,2} = \num{-0.1005 \pm 0.0020}$, $d_{1,2} = \num{-47.91 \pm 0.06}$, and $e_{1,2} = \num{-4.570 \pm 0.024}$, where errors are purely statistical at this stage. This result allows to evaluate the light-cone \ac{da} as given by \cref{eq:da-parametrization} including only $n=0$. The result can be seen on the \acs{rhs} of \cref{fig:da}.

\section*{Conclusions and outlook}

We apply \acl{sdf} to compute the \acl{da} of the $\Petac$ meson. Our lattice dataset consists of three $N_f=2$ \acs{cls} ensembles at a pion mass $m_{\Ppi} \sim \qty{270}{\mega\eV}$ and three different lattice spacings. We describe the light-cone \ac{da} with the help of a basis of Jacobi polynomials \cref{eq:da-parametrization}, and we compute in closed form the relation between the former and the \acl{pda} using the polynomials orthogonality properties. We take the continuum limit using the fit function \cref{eq:continuum-extrapolation-model}, which models the leading $\order{a}$ lattice artifacts, the high-twist contamination, and the small differences of the meson masses in the ensembles. Besides, we observe a large effect coming from the $\Petac$ mass, which points to the need to include nonzero quark masses in the matching kernel.
Finally, we extract the light-cone \ac{da} to first order in \cref{eq:da-parametrization}.

The current status of the project serves as a proof of concept on how to get \acp{da} on the continuum. Future work will focus on extending the analysis to several more \acs{cls} ensembles of similar characteristics. They will extend the Ioffe time interval, hopefully constrain more terms in the series \cref{eq:da-parametrization}, and probe the pion mass dependence. Besides, we recently got access to several $N_f=2+1+1$ ensemble at the physical pion mass, which will allow to include the missing sea-quark effects and eliminate the systematics from the unphysical pion masses. We are interested in recomputing the matching kernel in \cref{eq:matching}, this time keeping the quark masses nonzero, which should improve our Ansatz for the continuum limit.

\acknowledgments

The work by M.~T.~San Jos\'{e} is supported by Agence Nationale de la Recherche under the contract ANR-17-CE31-0019. The work by J.~M.~Morgado Ch\'{a}vez has been supported by P2IO LabEx (ANR-10-LABX-0038) in the framework of Investissements d’Avenir (ANR-11-IDEX-0003-01). This project was granted access to the HPC resources of TGCC (2021-A0100502271, 2022-A0120502271 and 2023-A0140502271) by GENCI. The authors thank Michael Fucilla, C\'{e}dric Mezrag, Lech Szymanowski and Samuel Wallon for valuable discussions.

\bibliographystyle{JHEP}
\bibliography{bib}

\providecommand{\href}[2]{#2}\begingroup\raggedright\begin{thebibliography}{10}

\bibitem{Lansberg:2019adr}
Lansberg, \emph{{New Observables in Inclusive Production of Quarkonia}},
  \href{https://doi.org/10.1016/j.physrep.2020.08.007}{\emph{Phys. Rept.}
  {\bfseries 889} (2020) 1}.

\bibitem{Han:2022rwq}
Han, Leibovich, Ma and Tan, \emph{{Higgs boson decay to charmonia via c-quark
  fragmentation}}, \href{https://doi.org/10.1007/JHEP08(2022)073}{\emph{JHEP}
  {\bfseries 08} (2022) 073}.

\bibitem{Radyushkin:2017cyf}
Radyushkin, \emph{{Quasi-parton distribution functions, momentum distributions,
  and pseudo-PDFs}},
  \href{https://doi.org/10.1103/PhysRevD.96.034025}{\emph{Phys. Rev. D}
  {\bfseries 96} (2017) 034025}.

\bibitem{Ji:2013dva}
Ji, \emph{{Parton Physics on a Euclidean Lattice}},
  \href{https://doi.org/10.1103/PhysRevLett.110.262002}{\emph{Phys. Rev. Lett.}
  {\bfseries 110} (2013) 262002}.

\bibitem{Orginos:2017kos}
Orginos, Radyushkin, Karpie and Zafeiropoulos, \emph{{Lattice QCD exploration
  of parton pseudo-distribution functions}},
  \href{https://doi.org/10.1103/PhysRevD.96.094503}{\emph{Phys. Rev. D}
  {\bfseries 96} (2017) 094503}.

\bibitem{Karpie:2018zaz}
Karpie, Orginos and Zafeiropoulos, \emph{{Moments of Ioffe time parton
  distribution functions from non-local matrix elements}},
  \href{https://doi.org/10.1007/JHEP11(2018)178}{\emph{JHEP} {\bfseries 11}
  (2018) 178}.

\bibitem{Radyushkin:2019owq}
Radyushkin, \emph{{GPDs and pseudodistributions}},
  \href{https://doi.org/10.1103/PhysRevD.100.116011}{\emph{Phys. Rev. D}
  {\bfseries 100} (2019) 116011}.

\bibitem{FlavourLatticeAveragingGroupFLAG:2021npn}
{\scshape FLAG} collaboration, \emph{{FLAG Review 2021}},
  \href{https://doi.org/10.1140/epjc/s10052-022-10536-1}{\emph{Eur. Phys. J. C}
  {\bfseries 82} (2022) 869}.

\bibitem{Karpie:2021pap}
{\scshape HadStruc} collaboration, \emph{{The continuum and leading twist
  limits of parton distribution functions in lattice QCD}},
  \href{https://doi.org/10.1007/JHEP11(2021)024}{\emph{JHEP} {\bfseries 11}
  (2021) 024}.

\bibitem{Belitsky:2005qn}
Belitsky and Radyushkin, \emph{{Unraveling hadron structure with GPDs}},
  \href{https://doi.org/10.1016/j.physrep.2005.06.002}{\emph{Phys. Rept.}
  {\bfseries 418} (2005) 1}.

\bibitem{Fritzsch:2012wq}
Fritzsch, Knechtli, Leder, Marinkovic, Schaefer, Sommer et~al., \emph{{The
  strange quark mass and Lambda parameter of two flavor QCD}},
  \href{https://doi.org/10.1016/j.nuclphysb.2012.07.026}{\emph{Nucl. Phys. B}
  {\bfseries 865} (2012) 397}.

\bibitem{Heitger:2013oaa}
Heitger, von Hippel, Schaefer and Virotta, \emph{{Charm quark mass and D-meson
  decay constants from two-flavour lattice QCD}},
  \href{https://doi.org/10.22323/1.187.0475}{\emph{PoS} {\bfseries LATTICE2013}
  (2014) 475}.

\bibitem{ParticleDataGroup:2022pth}
{\scshape PDG} collaboration, \emph{{Review of Particle Physics}},
  \href{https://doi.org/10.1093/ptep/ptac097}{\emph{PTEP} {\bfseries 2022}
  (2022) 083C01}.

\bibitem{doi:10.1137/0710036}
Golub and Pereyra, \emph{The differentiation of pseudo-inverses and nonlinear
  least squares problems whose variables separate},
  \href{https://doi.org/10.1137/0710036}{\emph{SIAM Journal on Numerical
  Analysis} {\bfseries 10} (1973) 413}.

\end{thebibliography}\endgroup

\end{document}